# Interpreting fluorescence detected 2D electronic spectroscopy

Oliver Kühn[1], Tomas Mancal[2], Tõnu Pullerits[3*]

1) Institute of Physics, University of Rostock, Albert Einstein Strasse 23-24, 18059 Rostock, Germany
2) Faculty of Mathematics and Physics, Charles University in Prague, Ke Karlovu 5, CZ-121 16 Prague 2, Czech Republic
3) Department of Chemical Physics and NanoLund, Lund University, P.O. Box 124, 22100 Lund, Sweden

*Corresponding author: tonu.pullerits@chemphys.lu.se

Electronic coherent multidimensional spectroscopy (CMDS) allows detailed studies of ultrafast dynamics in systems with highly congested complicated spectral bands.[1,2,3] A wide variety of experimental implementations have been developed[4] emphasising different strengths of the methodology. Originally, the 2D electronic spectroscopy (2DES) was based on phase-sensitive detection of a coherent photon-echo type signal generated by three laser pulses.[5,6] During the last decade four-pulse generated incoherent action detection has become popular. The concept was originally developed for fluorescence detected 2DES (F-2DES).[7] Later numerous other incoherent action detection methods have been used, e.g., based on photoelectrons,[8] photocurrents[9,10] and photo ions.[11]

Even though 2DES and F-2DES are analogous coherent techniques providing detailed information on ultrafast dynamics in two spectral dimensions, the methods have important differences and the measured spectra do not look the same.[12,13,14,15] While the "anatomy" of the 2DES has been studied in numerous articles[16,17,18] and is quite well understood, interpretation of the F-2DES spectra[19,20,21,22,23] has not reached similar maturity and is currently being debated in literature. In this Viewpoint we briefly summarize the various approaches to interpreting the F-2DES and present our view with the intension to resolve at least some of the existing controversial issues. We start from a brief summary of the two methods.

In 2DES the third order macroscopic polarization induced by three short laser pulses generates various coherent light fields in different phase-matched directions. For phase-sensitive detection the coherent field of a certain phase-matched direction is mixed with a fourth, so called local oscillator pulse and spectrally dispersed in a spectrometer. The Fourier transform of the delay between the first two pulses provides the excitation frequency axis (usually horizontal) and the spectrum of the coherent photon echo and free induction decay type signal fields give the detection axis (usually vertical).

In F-2DES four laser pulses bring the system to a fourth-order excited state population which is detected via the emitted fluorescence. For disentangling the detected signal into the components of interest, phase modulation of the pulses together with lock-in detection is used.[24,25] Alternatively, in phase cycling, a certain set of phase combinations of the pulses is applied thereby allowing to uniquely determine the signals of interest.[26]

Conceptually, there are three important differences between 2DES and F-2DES: 1) Since F-2DES has one more excitation pulse, there is an additional signal generating Liouville pathway as can be seen in Figure 1. 2) After the fourth pulse, fluorescence is usually emitted over long time period during which various processes can occur that influence the fluorescence yield and can mix two-pulse modulations.[27] 3) Since in F-2DES disentangling of the signals is achieved by phase modulation and corresponding Fourier filtering, no spatial summation (spatial Fourier filtering) for phase matching is needed which means that signals can be taken from samples smaller than the wavelength.[8] In addition, there is an important technical difference – while the nonresonant signal (from the solvent) can significantly distort the 2DES spectra during pulse overlap, fluorescence does not have any such nonresonant component. The latter is important not only for the double quantum coherence (DQC) signal[28] which is mainly generated during the pulse overlap[29] but also for the rephasing (R) and nonrephasing (NR) spectra at short population times comparable to the pulse length or shorter.

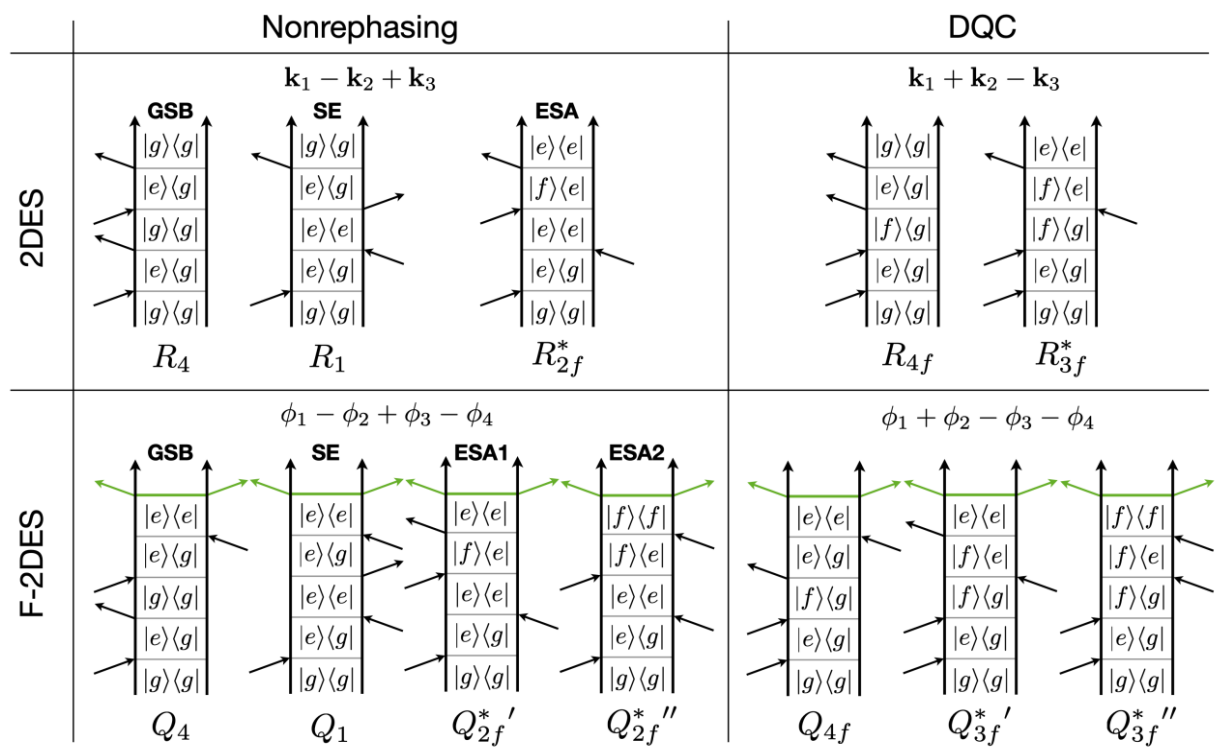

*Figure 1* Comparison of 2DES and F-2DES double-sided Feynman diagrams for a three-level system where |g>, |e> and |f> denote the ground-, the first and the doubly excited state, respectively. |f> can be a two-exciton state of a molecular aggregate, but also a higher excited state of a system like molecule or nanocrystal with energy about twice the energy of |e> that can be reached when |e> absorbs a photon. Rephasing pathways are not shown, they can be obtained by switching the time ordering of the first two interactions of the nonrephasing pathways. For 2DES we follow the usual response function notation, R, of ref [30]. The index f indicates that the doubly excited state |f> is involved. Different rephasing and nonrephasing pathways have an intuitive relation to experimentally observable ground state bleach (GSB), stimulated emission (SE) and excited state absorption (ESA) signals – the terminology widely used in pump-probe spectroscopy. In case of F-DES we use Q to denote the signal components as in ref [31]. The respective phase conditions are indicated. There is a general one-to-one analogy between R and Q pathways.[2] Owing to the fourth pulse, the F-2DES has two possibilities for ESA ending at either in an |e> or |f> population state. Since these two populations can give a different amplitude of the incoherent action signal (fluorescence) A, their contribution can be different. Though, in fluorescence detection many systems give just one photon from both ESA1 and ESA2. Since the overall sign of a diagram is determined by the number of incoming arrows from the right, EAS1 and ESA2 differ in sign such that the total ESA signal cancels out. Also in DQC an additional pathway appears in F-2DES and again, in many cases |e> and |f> populations give the same amount of emission leading to cancellation of the $Q_{3f}^{*'}$ and $Q_{3f}^{*''}$ pathways.

In the following we formulate a thought experiment based on a system of two well separated independent chromophores with excited electronic states $e_1$ and $e_2$ having different energies $E_1$ and $E_2$. The separation of the chromophores is assumed to be large enough (more than the wavelength of the emitted light) such that fluorescence of the individual systems can be independently measured. This large distance also implies that there is no interaction between the chromophores. Consequently, no annihilation can occur either. Further, we assume ideal detection where all emitted photons will be individually registered, see Fig. 2.

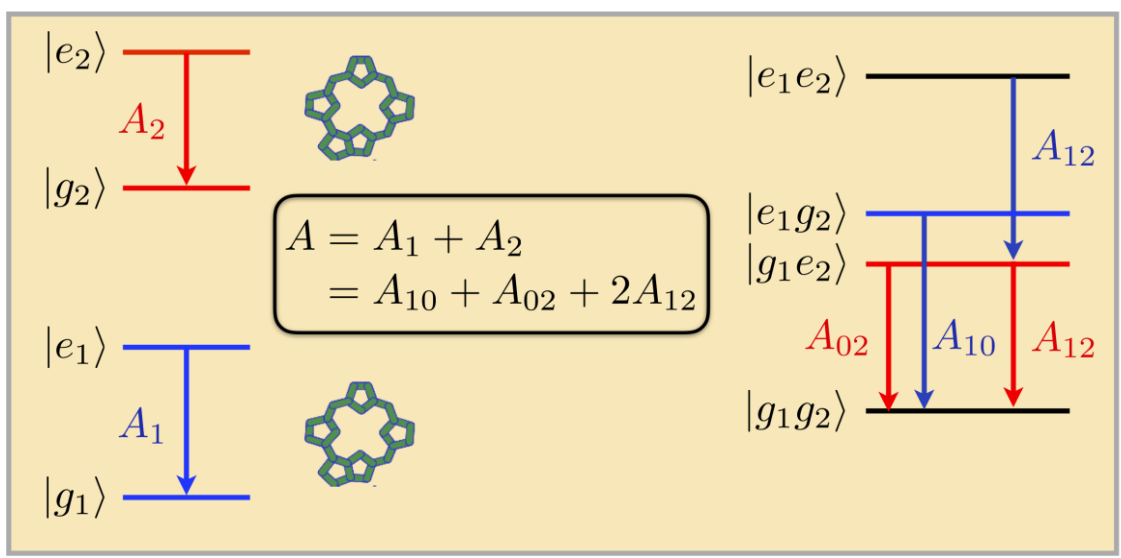

*Figure 2. Schematic illustration of the thought experiment where emission of two well separated systems both covered by the excitation beam can be independently monitored. The emission is described in one (left side) and two particle (right side) representation. See text for further discussion.*

In such an experiment we can define many different emission detection modes. We will be able to detect the total emission from the sample, $A$, one-particle emission from chromophore one, $A_1$, and from chromophore two, $A_2$, also the two-particle emission signals where both chromophores emit, $A_{12}$, and where one emits while the other does not, $A_{10}$ and $A_{02}$. We point out the important difference between one and two particle emissions like $A_1$ and $A_{10}$, respectively. In $A_1$ we count all emitted photons from the chromophore one no matter what the chromophore two does while in $A_{10}$ we only count the photons from one if chromophore two did not emit (was in ground state). If they both emitted, that would be $A_{12}$. Obviously $A = A_1 + A_2 = A_{10} + A_{02} + 2A_{12}$. The factor two takes into account that the doubly excited state contributes two photons to the detected emission.

We can, in principle, detect $A_1$ and $A_2$ individually. In F-2DES they would contribute to their corresponding diagonal peaks at energies $E_1$ and $E_2$. Of course, by detecting the total emission $A$, the same diagonal peaks would appear and we would not have any cross peaks. Interestingly, if we were to use any of the two-particle detection modes, the cross peaks would be expected. The cross peaks from $A_{12}$ have analogous origin as the many-particle signals discussed by Mukamel.[32] Cross peaks from $A_{10}$ and $A_{02}$ may feel counter-intuitive at first glance. Here one has to realize that in the spirit of the Feynman diagrams, phase modulation is effective for both excited and nonexcited states. This means that even though in, for example, $A_{10}$, the emission comes only from the first chromophore, the joint probability $A_{10}$ also takes into account that the second chromophore has to be in the ground state. Both population probabilities are modulated. $A_{12}$, $A_{10}$ and $A_{02}$ directly correspond to a set of Feynman diagrams which further explain the origin of the modulation. In Figure 3 we present the diagrams both in one- (lower) and two- particle (upper) representation. It turns out that the singly ($A_{10}$ and $A_{02}$) and doubly ($A_{12}$) excited two-particle signals of cross peaks have opposite signs and exactly cancel out if detected together. In terms of modulation of the corresponding populations, the opposite signs of the signal pathways correspond to the opposite phase (phase shift of π) of population modulation which leads to the cancellation.[22] This reveals the physics of these cancellations.

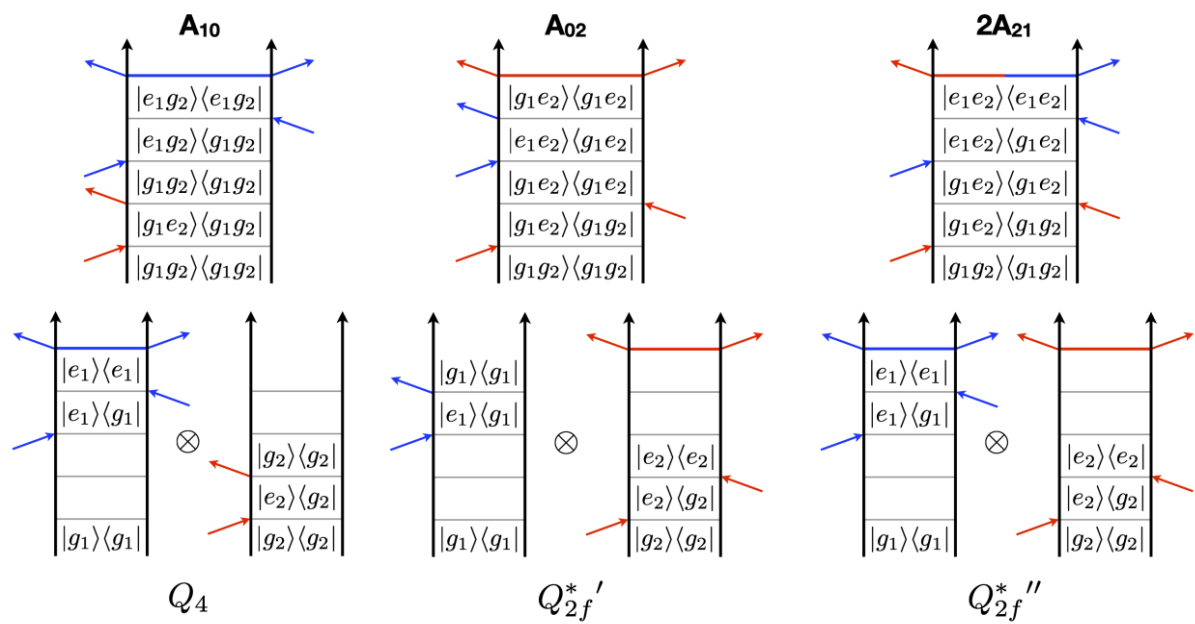

*Figure 3.* Two-particle emissions which contribute to the lower cross peak at energy ($E_2$, $E_1$) represented via nonrephasing double-sided Feynman diagrams. In the lower row the general two-particle states are broken down to explicit diagrams of the individual chromophores. For the colour code see Fig. 2.

Let us now assume a weak interaction between the chromophores. The interaction would allow excitation energy transfer between the chromophores. Such transfer can occur even in case where both chromophores are excited leading to a population of a highly excited state at energy equal to sum of the two initial excitations. The highly excited state of a chromophore will rapidly relax through internal conversion back to the lowest excited state. The process as a whole is called excitation annihilation – one excitation out of two disappears. Consequently, the two-particle signal component $A_{12}$ would give us less than two photons and the cancellation would not be perfect. This will lead to appearance of the cross peaks in F-2DES experiment. The strength of the cross peaks would depend on the efficiency of annihilation and one can argue that the cross peaks can be taken as the signature of annihilation.[20,21]

In the extreme case of close to 100% annihilation a different kind of cancelation occurs. In this case the ESA2 pathway which corresponds to the double excitation of the complex will only give one photon exactly cancelling out ESA1, cf. Fig. 1. This means that F-2DES in case of perfect annihilation would be ESA free.[13,19] As a side line, we point out that the conclusion[23] that the annihilation does not contribute to the nonlinear signal detected in F-2DES can be directly related to the exact cancellation of these two signal components. In general, the ESA signal contains valuable information about many-body effects in excited states of the aggregate. Unfortunately, usually ESA is heavily overlapping with GSB and SE and frequently involves doubly excited monomeric states making it hard to analyse. The overlap also means that in a usual pump-probe or 2DES experiment GSB and SE are not "cleanly" available either. In F-2DES, in case of perfect annihilation, one has access to undisturbed GSB and SE signal at the cross peaks which, the same way as linear absorption, can be used to evaluate collective phenomena like exciton coherence domain. However, it has been pointed out that the cross peak behaviour is not trivial.[19,20] For example, the cross peak amplitude can increase while the coupling strength is decreasing. Still, if structure is known, the exciton mixing angle can be evaluated.

Alternatively, instead of focusing on the cross peak amplitude/strength, one could construct a signal analogous to the so-called two colour photon echo peak-shift experiment. The method has been used to assess the exciton mixing angle of a molecular dimer, without knowing its structure.[33] Ideally the method would be based on the GSB part of the signal only, and it was shown to provide reasonable estimates despite the presence of ESA. In the F-2DES, however, the GSB only signal could be isolated and analysed as in[33] to yield an even more precise estimate of the mixing angle.

A prototypical system where both 2DES and F-2DES measurements have been reported and show principal differences is the peripheral light harvesting complex of photosynthetic purple bacteria, the so called LH2. The complex consists of two rings of bacteriochlorophyll *a* (BChl) molecules called B800 and B850 according to their near infrared absorption maxima.[34] The B850 ring is densely packed with BChl nearest neighbour excitonic couplings around 300 cm$^{-1}$.[35] The coupling within B850 is known to cause significant excitonic effects leading to delocalization of the electronically excited states.[36,37] The coupling between B800 BChls and between B800 and B850 is about 20 cm$^{-1}$. 2DES experiments[38] and theoretical considerations[39,40] have suggested that, despite of rather weak interaction, delocalization effects occur within B800. Delocalization between B800 and B850 BChl molecules[41] and within B850[42] has also been invoked to explain the interband transfer.

Recent F-2DES experiments on LH2 from *Rps. acidophila* and *Rps. palustris* reveal pronounced cross peaks at early population times.[12,13] Corresponding cross peaks in short population time 2DES are weak if not absent.[15,43] The two colour pump-probe experiments with B800 excitation and B850 detection are consistent with 2DES showing initially negligible bleach signal.[44] These differences are well explained by the additional ESA2 pathway which, in case of efficient annihilation, as in LH2,[45,46] cancels the conventional ESA1 revealing clear bleach + stimulated emission signal.[13] It is commonly agreed that the cross peaks in 2DES are signatures of correlations and exciton delocalization of the involved states.[47] For the F-2DES there is no such consensus. On one hand, the undisturbed GSB + SE in F-2DES can allow more quantitative analyses of the cross peaks in terms of correlations compared to the case where the signal is analysed as GSB + SE - ESA1 where the ESA1 is typically not as well determined and described as the other two.[13,36] On the other hand, the ESA2 – ESA1 cancellation (which reveals the cross peaks) takes place due to the exciton-exciton annihilation and thereby the cross peaks can be taken as signatures of the annihilation.[20,21] In case of individual LH2s where perfect annihilation is a well-established fact, the latter aspect is less of interest while the potential of the method for evaluating delocalization is more relevant. However, connectivity in large antenna complexes[48] can be possibly investigated via existence of annihilation witnessed by the F-2DES cross peaks.

Finally, we point out that our considerations here may be relevant for the discussion of the two-particle DQC signal observed in dilute gases.[49] These experiments have been earlier discussed in terms of a two-particle correlated signal[32] but also as a result of a very weak dipole-dipole interaction.[50] Our discussion together with careful analyses of what is actually measured in such experiments may be able to distinguish between different possibilities.

In summary, the two main interpretation lines of the cross peaks in F-DES spectra, delocalization and annihilation, correspond to different annihilation efficiency. In the case of very efficient annihilation, total cancellation of the ESA signal can allow a more quantitative estimates of exciton delocalization and correlation effects based on the cross peak signals. In the case of weak connectivity in extended complexes the early time cross peaks can also be taken as direct evidence of the annihilation.

**Acknowledgments**

This work was supported by the Swedish Research Council VR, Swedish Energy Agency, KAW and Laserlab-Europe EU-H2020 grant 654148. T.P. thanks the Mare Balticum program for funding his stay at Rostock University. T.M. acknowledges funding by the Czech Science Foundation (GAČR) through the grant no. 17-22160S.